\documentclass{bioinfo}
\copyrightyear{2018}
\pubyear{2018}

\usepackage{graphicx}
\usepackage{hyperref}
\usepackage{url}
\usepackage{amsmath}
\usepackage[ruled,vlined]{algorithm2e}

\SetCommentSty{mycommfont}
\SetKwComment{Comment}{$\triangleright$\ }{}

\usepackage{natbib}
\DeclareMathOperator*{\argmax}{argmax}

\begin{document}
\firstpage{1}

\title[Aligning nucleotide sequences with minimap2]{Minimap2: pairwise alignment for nucleotide sequences}
\author[Li]{Heng Li}
\address{Broad Institute, 415 Main Street, Cambridge, MA 02142, USA}

\maketitle

\begin{abstract}

\section{Motivation:} Recent advances in sequencing technologies promise
ultra-long reads of $\sim$100 kilo bases (kb) in average, full-length mRNA or
cDNA reads in high throughput and genomic contigs over 100 mega bases (Mb) in
length. Existing alignment programs are unable or inefficient to process such data
at scale, which presses for the development of new alignment algorithms.

\section{Results:} Minimap2 is a general-purpose alignment program to map DNA or long
mRNA sequences against a large reference database. It works with accurate short
reads of $\ge$100bp in length, $\ge$1kb genomic reads at error rate $\sim$15\%,
full-length noisy Direct RNA or cDNA reads, and assembly contigs or closely
related full chromosomes of hundreds of megabases in length. Minimap2 does
split-read alignment, employs concave gap cost for long insertions and
deletions (INDELs) and introduces new heuristics to reduce spurious alignments.
It is 3--4 times as fast as mainstream short-read mappers at comparable
accuracy, and is $\ge$30 times faster than long-read genomic or cDNA
mappers at higher accuracy, surpassing most aligners specialized in one type of
alignment.

\section{Availability and implementation:}
\href{https://github.com/lh3/minimap2}{https://github.com/lh3/minimap2}

\section{Contact:} hengli@broadinstitute.org
\end{abstract}

\section{Introduction}

Single Molecule Real-Time (SMRT) sequencing technology and Oxford Nanopore
technologies (ONT) produce reads over 10kbp in length at an error rate
$\sim$15\%. Several aligners have been developed for such
data~\citep{Chaisson:2012aa,Li:2013aa,Liu:2016ab,Sovic:2016aa,Liu:2017aa,Lin:2017aa,Sedlazeck169557}.
Most of them were five times as slow as mainstream short-read
aligners~\citep{Langmead:2012fk,Li:2013aa} in terms of the number of bases
mapped per second. We speculated there could be substantial room for speedup on
the thought that 10kb long sequences should be easier to map than 100bp reads
because we can more effectively skip repetitive regions, which are often the
bottleneck of short-read alignment. We confirmed our speculation by achieving
approximate mapping 50 times faster than BWA-MEM~\citep{Li:2016aa}.
\citet{Suzuki:2018aa} extended our work with a fast and novel algorithm on
generating base-level alignment, which in turn inspired us to develop minimap2
with added functionality.

Both SMRT and ONT have been applied to the sequencing of spliced mRNAs (RNA-seq). While
traditional mRNA aligners work~\citep{Wu:2005vn,Iwata:2012aa}, they are not
optimized for long noisy sequence reads and are tens of times slower than
dedicated long-read aligners. When developing minimap2 initially for aligning
genomic DNA only, we realized minor modifications could enable the base
algorithm to map mRNAs as well. Minimap2 becomes a first RNA-seq aligner
specifically designed for long noisy reads. We have also extended the original
algorithm to map short reads at a speed faster than several mainstream
short-read mappers.

In this article, we will describe the minimap2 algorithm and its applications
to different types of input sequences. We will evaluate the performance and
accuracy of minimap2 on several simulated and real data sets and demonstrate
the versatility of minimap2.

\begin{methods}
\section{Methods}

Minimap2 follows a typical seed-chain-align procedure as is used by most
full-genome aligners. It collects minimizers~\citep{Roberts:2004fv} of the
reference sequences and indexes them in a hash table, with the key being the
hash of a minimizer and the value being a list of locations of the minimizer
copies. Then for each query
sequence, minimap2 takes query minimizers as \emph{seeds}, finds exact matches
(i.e. \emph{anchors}) to the reference, and identifies sets of colinear anchors as
\emph{chains}. If base-level alignment is requested, minimap2 applies dynamic
programming (DP) to extend from the ends of chains and to close
regions between adjacent anchors in chains.

Minimap2 uses indexing and seeding algorithms similar to
minimap~\citep{Li:2016aa}, and furthers the predecessor with more accurate
chaining, the ability to produce base-level alignment and the support of
spliced alignment.

\subsection{Chaining}

\subsubsection{Chaining}
An \emph{anchor} is a 3-tuple $(x,y,w)$, indicating interval $[x-w+1,x]$ on the
reference matching interval $[y-w+1,y]$ on the query. Given a list of anchors
sorted by ending reference position $x$, let $f(i)$ be the maximal chaining
score up to the $i$-th anchor in the list. $f(i)$ can be calculated with
dynamic programming:
\begin{equation}\label{eq:chain}
f(i)=\max\big\{\max_{i>j\ge 1} \{ f(j)+\alpha(j,i)-\beta(j,i) \},w_i\big\}
\end{equation}
where $\alpha(j,i)=\min\big\{\min\{y_i-y_j,x_i-x_j\},w_i\big\}$ is the number of
matching bases between the two anchors. $\beta(j,i)>0$ is the gap cost. It
equals $\infty$ if $y_j\ge y_i$ or $\max\{y_i-y_j,x_i-x_j\}>G$ (i.e. the
distance between two anchors is too large); otherwise
\begin{equation}\label{eq:chain-gap}
\beta(j,i)=\gamma_c\big((y_i-y_j)-(x_i-x_j)\big)
\end{equation}
In implementation, a gap of length $l$ costs
\[
\gamma_c(l)=\left\{\begin{array}{ll}
0.01\cdot \bar{w}\cdot|l|+0.5\log_2|l| & (l\not=0) \\
0 & (l=0)
\end{array}\right.
\]
where $\bar{w}$ is the average seed length. For $N$ anchors, directly computing all $f(\cdot)$ with
Eq.~(\ref{eq:chain}) takes $O(N^2)$ time. Although theoretically faster
chaining algorithms exist~\citep{Abouelhoda:2005aa}, they
are inapplicable to generic gap cost, complex to implement and usually
associated with a large constant. We introduced a simple heuristic to
accelerate chaining.

We note that if anchor $i$ is chained to $j$, chaining $i$ to a predecessor
of $j$ is likely to yield a lower score. When evaluating Eq.~(\ref{eq:chain}),
we start from anchor $i-1$ and stop the process if we cannot find a better
score after up to $h$ iterations. This approach reduces the average time to
$O(hN)$. In practice, we can almost always find the optimal chain with
$h=50$; even if the heuristic fails, the optimal chain is often close.

\subsubsection{Backtracking}
Let $P(i)$ be the index of the best predecessor of anchor $i$. It equals 0 if
$f(i)=w_i$ or $\argmax_j\{f(j)+\alpha(j,i)-\beta(j,i)\}$ otherwise. For each
anchor $i$ in the descending order of $f(i)$, we apply $P(\cdot)$ repeatedly to
find its predecessor and mark each visited $i$ as `used', until $P(i)=0$ or we
reach an already `used' $i$. This way we find all chains with no anchors used
in more than one chains.

\subsubsection{Identifying primary chains}\label{sec:primary}
In the absence of copy number changes, each query segment should not be mapped
to two places in the reference. However, chains found at the previous step may
have significant or complete overlaps due to repeats in the reference~\citep{Li:2010fk}.
Minimap2 used the following procedure to identify \emph{primary chains} that do
not greatly overlap on the query.

Let $Q$ be an empty set initially. For each
chain from the best to the worst according to their chaining scores: if on the
query, the chain overlaps with a chain in $Q$ by 50\% or higher percentage of
the shorter chain, mark the chain as secondary to the chain in $Q$; otherwise,
add the chain to $Q$. In the end, $Q$ contains all the primary chains. We did
not choose a more sophisticated data structure (e.g. range tree or k-d tree)
because this step is not the performance bottleneck.

For each primary chain, minimap2 estimates its mapping quality with an
empirical formula:
\[
{\rm mapQ}=40\cdot (1-f_2/f_1)\cdot\min\{1,m/10\}\cdot\log f_1
\]
where $\log$ denotes natural logarithm, $m$ is the number of anchors on the primary chain, $f_1$ is the chaining
score, and $f_2\le f_1$ is the score of the best chain that is secondary to the
primary chain. Intuitively, a chain is assigned to a higher mapping quality if
it is long and its best secondary chain is weak.

\subsubsection{Estimating per-base sequence divergence}
Suppose a query sequence harbors $n$ seeds of length $k$, $m$ of which are
present in a chain. We want to estimate the sequence divergence $\epsilon$
between the query and the reference sequences in the chain. This is useful
when base-level alignment is too expensive to perform.

If we model substitutions with a homogeneous Poisson process along the query
sequence, the probablity of seeing $k$ consecutive bases without substitutions
is $e^{-k\epsilon}$. On the assumption that all $k$-mers are independent of
each other, the likelihood function of $\epsilon$ is
\[
\mathcal{L}(\epsilon|n,m,k)=e^{-m\cdot k\epsilon}(1-e^{-k\epsilon})^{n-m}
\]
The maximum likelihood estimate of $\epsilon$ is
\[
\hat{\epsilon}=\frac{1}{k}\log\frac{n}{m}
\]
In reality, sequencing errors are sometimes clustered and $k$-mers are not
independent of each other, especially when we take minimizers as seeds. These
violate the assumptions in the derivation above. As a result, $\hat{\epsilon}$
is only approximate and can be biased. It also ignores long deletions from the
reference sequence. In practice, fortunately, $\hat{\epsilon}$ is often close
to and strongly correlated with the sequence divergence estimated from
base-level alignments. On the several datasets used in
Section~\ref{sec:long-genomic}, the Spearman correlation coefficient is around
$0.9$.

\subsubsection{Indexing with homopolymer compressed $k$-mers}
SmartDenovo
(\href{https://github.com/ruanjue/smartdenovo}{https://github.com/ruanjue/smartdenovo};
J. Ruan, personal communication) indexes reads with homopolymer-compressed (HPC)
$k$-mers and finds the strategy improves overlap sensitivity for SMRT reads.
Minimap2 adopts the same heuristic.

The HPC string of a string $s$, denoted by ${\rm HPC}(s)$, is constructed by
contracting homopolymers in $s$ to a single base.  An HPC $k$-mer of $s$ is a
$k$-long substring of ${\rm HPC}(s)$. For example, suppose $s={\tt GGATTTTCCA}$,
${\rm HPC}(s)={\tt GATCA}$ and the first HPC 4-mer is ${\tt GATC}$.

To demonstrate the effectiveness of HPC $k$-mers, we performed read overlapping
for the example {\it E. coli} SMRT reads from PBcR~\citep{Berlin:2015xy}, using
different types of $k$-mers. With normal 15bp minimizers per 5bp window,
minimap2 finds 90.9\% of $\ge$2kb overlaps inferred from the read-to-reference
alignment. With HPC 19-mers per 5bp window, minimap2 finds 97.4\% of overlaps. It achieves this
higher sensitivity by indexing 1/3 fewer minimizers, which further helps
performance. HPC-based indexing reduces the sensitivity for current ONT reads, though.

\subsection{Aligning genomic DNA}\label{sec:genomic}

\subsubsection{Alignment with 2-piece affine gap cost}

Minimap2 performs DP-based global alignment between adjacent anchors in a
chain. It uses a 2-piece affine gap cost~\citep{Gotoh:1990aa}:
\begin{equation}\label{eq:2-piece}
\gamma_a(l)=\min\{q+|l|\cdot e,\tilde{q}+|l|\cdot\tilde{e}\}
\end{equation}
Without losing generality, we always assume $q+e<\tilde{q}+\tilde{e}$.
On the condition that $e>\tilde{e}$, it applies cost $q+|l|\cdot e$ to gaps
shorter than $\lceil(\tilde{q}-q)/(e-\tilde{e})\rceil$ and applies
$\tilde{q}+|l|\cdot\tilde{e}$ to longer gaps.  This scheme helps to recover
longer insertions and deletions~(INDELs).

The equation to compute the optimal alignment under $\gamma_a(\cdot)$ is
\begin{equation}\label{eq:ae86}
\left\{\begin{array}{l}
H_{ij} = \max\{H_{i-1,j-1}+s(i,j),E_{ij},F_{ij},\tilde{E}_{ij},\tilde{F}_{ij}\}\\
E_{i+1,j}= \max\{H_{ij}-q,E_{ij}\}-e\\
F_{i,j+1}= \max\{H_{ij}-q,F_{ij}\}-e\\
\tilde{E}_{i+1,j}= \max\{H_{ij}-\tilde{q},\tilde{E}_{ij}\}-\tilde{e}\\
\tilde{F}_{i,j+1}= \max\{H_{ij}-\tilde{q},\tilde{F}_{ij}\}-\tilde{e}
\end{array}\right.
\end{equation}
where $s(i,j)$ is the score between the $i$-th reference base and $j$-th query
base. Eq.~(\ref{eq:ae86}) is a natural extension to the equation under affine
gap cost~\citep{Gotoh:1982aa,Altschul:1986aa}.

\subsubsection{The Suzuki-Kasahara formulation}

When we allow gaps longer than several hundred base pairs, nucleotide-level
alignment is much slower than chaining. SSE acceleration is critical to the
performance of minimap2. Traditional SSE implementations~\citep{Farrar:2007hs}
based on Eq.~(\ref{eq:ae86}) can achieve 16-way parallelization for short
sequences, but only 4-way parallelization when the peak alignment score reaches
32767. Long sequence alignment may exceed this threshold. Inspired by
\citet{Wu:1996aa} and the following work, \citet{Suzuki:2018aa} proposed a
difference-based formulation that lifted this limitation.
In case of 2-piece gap cost, define
\[
\left\{\begin{array}{ll}
u_{ij}\triangleq H_{ij}-H_{i-1,j} & v_{ij}\triangleq H_{ij}-H_{i,j-1} \\
x_{ij}\triangleq E_{i+1,j}-H_{ij} & \tilde{x}_{ij}\triangleq \tilde{E}_{i+1,j}-H_{ij} \\
y_{ij}\triangleq F_{i,j+1}-H_{ij} & \tilde{y}_{ij}\triangleq \tilde{F}_{i,j+1}-H_{ij}
\end{array}\right.
\]
We can transform Eq.~(\ref{eq:ae86}) to
\begin{equation}\label{eq:suzuki}
\left\{\begin{array}{lll}
z_{ij}&=&\max\{s(i,j),x_{i-1,j}+v_{i-1,j},y_{i,j-1}+u_{i,j-1},\\
&&\tilde{x}_{i-1,j}+v_{i-1,j},\tilde{y}_{i,j-1}+u_{i,j-1}\}\\
u_{ij}&=&z_{ij}-v_{i-1,j}\\
v_{ij}&=&z_{ij}-u_{i,j-1}\\
x_{ij}&=&\max\{0,x_{i-1,j}+v_{i-1,j}-z_{ij}+q\}-q-e\\
y_{ij}&=&\max\{0,y_{i,j-1}+u_{i,j-1}-z_{ij}+q\}-q-e\\
\tilde{x}_{ij}&=&\max\{0,\tilde{x}_{i-1,j}+v_{i-1,j}-z_{ij}+\tilde{q}\}-\tilde{q}-\tilde{e}\\
\tilde{y}_{ij}&=&\max\{0,\tilde{y}_{i,j-1}+u_{i,j-1}-z_{ij}+\tilde{q}\}-\tilde{q}-\tilde{e}
\end{array}\right.
\end{equation}
where $z_{ij}$ is a temporary variable that does not need to be stored.

An important property of Eq.~(\ref{eq:suzuki}) is that all values are bounded
by scoring parameters. To see that,
\[
x_{ij}=E_{i+1,j}-H_{ij}=\max\{-q,E_{ij}-H_{ij}\}-e
\]
With $E_{ij}\le H_{ij}$, we have
\[
-q-e\le x_{ij}\le\max\{-q,0\}-e=-e
\]
and similar inequations for $y_{ij}$, $\tilde{x}_{ij}$ and $\tilde{y}_{ij}$.
In addition,
\[
u_{ij}=z_{ij}-v_{i-1,j}\ge\max\{x_{i-1,j},\tilde{x}_{i-1,j}\}\ge-q-e
\]
As the maximum value of $z_{ij}=H_{ij}-H_{i-1,j-1}$ is $M$, the maximal
matching score, we can derive
\[
u_{ij}\le M-v_{i-1,j}\le M+q+e
\]
In conclusion, in Eq.~(\ref{eq:suzuki}), $x$ and $y$ are bounded by $[-q-e,-e]$,
$\tilde{x}$ and $\tilde{y}$ by $[-\tilde{q}-\tilde{e},-\tilde{e}]$, and $u$ and
$v$ by $[-q-e,M+q+e]$. When $-128\le-q-e<M+q+e\le127$, each of them can be stored as
a 8-bit integer. This enables 16-way SSE vectorization regardless of the peak
score of the alignment.

For a more efficient SSE implementation, we transform the row-column coordinate
to the diagonal-antidiagonal coordinate by letting $r\gets i+j$ and $t\gets i$.
Eq.~(\ref{eq:suzuki}) becomes:
\begin{equation*}
\left\{\begin{array}{lll}
z_{rt}&=&\max\{s(t,r-t),x_{r-1,t-1}+v_{r-1,t-1},y_{r-1,t}\\
&&+u_{r-1,t},\tilde{x}_{r-1,t-1}+v_{r-1,t-1},\tilde{y}_{r-1,t}+u_{r-1,t}\}\\
u_{rt}&=&z_{rt}-v_{r-1,t-1}\\
v_{rt}&=&z_{rt}-u_{r-1,t}\\
x_{rt}&=&\max\{0,x_{r-1,t-1}+v_{r-1,t-1}-z_{rt}+q\}-q-e\\
y_{rt}&=&\max\{0,y_{r-1,t}+u_{r-1,t}-z_{rt}+q\}-q-e\\
\tilde{x}_{rt}&=&\max\{0,\tilde{x}_{r-1,t-1}+v_{r-1,t-1}-z_{rt}+\tilde{q}\}-\tilde{q}-\tilde{e}\\
\tilde{y}_{rt}&=&\max\{0,\tilde{y}_{r-1,t}+u_{r-1,t}-z_{rt}+\tilde{q}\}-\tilde{q}-\tilde{e}
\end{array}\right.
\end{equation*}
In this formulation, cells with the same diagonal index $r$ are independent of
each other. This allows us to fully vectorize the computation of all cells on
the same anti-diagonal in one inner loop. It also simplifies banded alignment (500bp band width by default),
which would be difficult with striped vectorization~\citep{Farrar:2007hs}.

On the condition that $q+e<\tilde{q}+\tilde{e}$ and $e>\tilde{e}$, the initial
values in the diagonal-antidiagonal formuation are
\[
\left\{\begin{array}{l}
x_{r-1,-1}=y_{r-1,r}=-q-e\\
\tilde{x}_{r-1,-1}=\tilde{y}_{r-1,r}=-\tilde{q}-\tilde{e}\\
u_{r-1,r}=v_{r-1,-1}=\eta(r)\\
\end{array}\right.
\]
where
\[
\eta(r)=\left\{\begin{array}{ll}
-q-e & (r=0) \\
-e & (r<\lceil\frac{\tilde{q}-q}{e-\tilde{e}}-1\rceil) \\
r\cdot(e-\tilde{e})-(\tilde{q}-q)-\tilde{e} & (r=\lceil\frac{\tilde{q}-q}{e-\tilde{e}}-1\rceil) \\
-\tilde{e} & (r>\lceil\frac{\tilde{q}-q}{e-\tilde{e}}-1\rceil)
\end{array}\right.
\]
These can be derived from the initial values for Eq.~(\ref{eq:ae86}).

When performing global alignment, we do not need to compute $H_{rt}$ in each cell.
We use 16-way vectorization throughout the alignment process. When extending
alignments from ends of chains, we need to find the cell $(r,t)$ where $H_{rt}$
reaches the maximum. We resort to 4-way vectorization to compute
$H_{rt}=H_{r-1,t}+u_{rt}$. Because this computation is simple,
Eq.~(\ref{eq:suzuki}) is still the dominant performance bottleneck.

In practice, our 16-way vectorized implementation of global alignment is three
times as fast as Parasail's 4-way vectorization~\citep{Daily:2016aa}.  Without
banding, our implementation is slower than Edlib~\citep{Sosic:2017aa}, but with
a 1000bp band, it is considerably faster. When performing global alignment
between anchors, we expect the alignment to stay close to the diagonal of the
DP matrix. Banding is applicable most of the time.

\subsubsection{The Z-drop heuristic}

With global alignment, minimap2 may force to align unrelated sequences between
two adjacent anchors. To avoid such an artifact, we compute accumulative
alignment score along the alignment path and break the alignment where the
score drops too fast in the diagonal direction. More precisely, let $S(i,j)$ be
the alignment score along the alignment path ending at cell $(i,j)$ in the DP
matrix. We break the alignment if there exist $(i',j')$ and $(i,j)$, $i'<i$ and
$j'<j$, such that
\[
S(i',j')-S(i,j)>Z+e\cdot|(i-i')-(j-j')|
\]
where $e$ is the gap extension cost and $Z$ is an arbitrary threshold.
This strategy is first used in BWA-MEM. It is similar to X-drop employed in
BLAST~\citep{Altschul:1997vn}, but unlike X-drop, it would not break the
alignment in the presence of a single long gap.

When minimap2 breaks a global alignment between two anchors, it performs local
alignment between the two subsequences involved in the global alignment, but
this time with the one subsequence reverse complemented. This additional
alignment step may identify short inversions that are missed during chaining.

\subsubsection{Filtering out misplaced anchors}
Due to sequencing errors and local homology, some anchors in a chain may be
wrong. If we blindly align regions between two misplaced anchors, we will
produce a suboptimal alignment. To reduce this artifact, we filter out
anchors that lead to a $>$10bp insertion and a $>$10bp deletion at the same
time, and filter out terminal anchors that lead to a long gap towards the ends
of a chain. These heuristics greatly alleviate the issues with misplaced
anchors, but they are unable to fix all such errors. Local misalignment is a
limitation of minimap2 which we hope to address in future.

\subsection{Aligning spliced sequences}

The algorithm described above can be adapted to spliced alignment. In this
mode, the chaining gap cost distinguishes insertions to and deletions from the
reference: $\gamma_c(l)$ in Eq.~(\ref{eq:chain-gap}) takes the form of
\[
\gamma_c(l)=\left\{\begin{array}{ll}
0.01\cdot\bar{w}\cdot l+0.5\log_2 l & (l>0) \\
\min\{0.01\cdot\bar{w}\cdot|l|,\log_2|l|\} & (l<0)
\end{array}\right.
\]
Similarly, the gap cost function used for DP-based alignment is changed to
\[
\gamma_a(l)=\left\{\begin{array}{ll}
q+l\cdot e & (l>0) \\
\min\{q+|l|\cdot e,\tilde{q}\} & (l<0)
\end{array}\right.
\]
In alignment, a deletion no shorter than $\lceil(\tilde{q}-q)/e\rceil$ is
regarded as an intron, which pays no cost to gap extensions.

To pinpoint precise splicing junctions, minimap2 introduces reference-dependent
cost to penalize non-canonical splicing:
\begin{equation}\label{eq:splice}
\left\{\begin{array}{l}
H_{ij} = \max\{H_{i-1,j-1}+s(i,j),E_{ij},F_{ij},\tilde{E}_{ij}-a(i)\}\\
E_{i+1,j}= \max\{H_{ij}-q,E_{ij}\}-e\\
F_{i,j+1}= \max\{H_{ij}-q,F_{ij}\}-e\\
\tilde{E}_{i+1,j}= \max\{H_{ij}-d(i)-\tilde{q},\tilde{E}_{ij}\}\\
\end{array}\right.
\end{equation}
Let $T$ be the reference sequence. $d(i)$ is computed as
\[d(i)=\left\{\begin{array}{ll}
0 & \mbox{if $T[i+1,i+3]$ is ${\tt GTA}$ or ${\tt GTG}$} \\
p/2 & \mbox{if $T[i+1,i+3]$ is ${\tt GTC}$ or ${\tt GTT}$} \\
p & \mbox{otherwise}
\end{array}\right.\]
where $T[i,j]$ extracts a substring of $T$ between $i$ and $j$ inclusively.
$d(i)$ penalizes non-canonical donor sites with $p$ and less frequent Eukaryotic
splicing signal ${\tt GT[C/T]}$ with $p/2$~\citep{Irimia:2008aa}. Similarly,
\[a(i)=\left\{\begin{array}{ll}
0 & \mbox{if $T[i-2,i]$ is ${\tt CAG}$ or ${\tt TAG}$} \\
p/2 & \mbox{if $T[i-2,i]$ is ${\tt AAG}$ or ${\tt GAG}$} \\
p & \mbox{otherwise}
\end{array}\right.\]
models the acceptor signal. Eq.~(\ref{eq:splice}) is close to an equation in
\citet{Zhang:2006aa} except that we allow insertions immediately followed by
deletions and vice versa; in addition, we use the Suzuki-Kasahara diagonal
formulation in actual implementation.

If RNA-seq reads are not sequenced from stranded libraries, the read strand
relative to the underlying transcript is unknown. By default, minimap2 aligns
each chain twice, first assuming ${\tt GT}$--${\tt AG}$ as the splicing signal
and then assuming ${\tt CT}$--${\tt AC}$, the reverse complement of ${\tt
GT}$--${\tt AG}$, as the splicing signal. The alignment with a higher score is
taken as the final alignment. This procedure also infers the relative strand of
reads that span canonical splicing sites.

In the spliced alignment mode, minimap2 further increases the density of
minimizers and disables banded alignment. Together with the two-round DP-based
alignment, spliced alignment is several times slower than genomic DNA
alignment.

\subsection{Aligning short paired-end reads}

During chaining, minimap2 takes a pair of reads as one fragment with a gap of
unknown length in the middle. It applies a normal gap cost between seeds on the
same read but is a more permissive gap cost between seeds on different reads.
More precisely, the gap cost during chaining is ($l\not=0$):
\[
\gamma_c(l)=\left\{\begin{array}{ll}
0.01\cdot\bar{w}\cdot |l|+0.5\log_2 |l| & \mbox{if two seeds on the same read} \\
\min\{0.01\cdot\bar{w}\cdot|l|,\log_2|l|\} & \mbox{otherwise}
\end{array}\right.
\]
After identifying primary chains (Section~\ref{sec:primary}), we split each
fragment chain into two read chains and perform alignment for each read as in
Section~\ref{sec:genomic}.  Finally, we pair hits of each read end to find
consistent paired-end alignments.

\end{methods}

\section{Results}

Minimap2 is implemented in the C programming language and comes with APIs in
both C and Python. It is distributed under the MIT license, free to both
commercial and academic uses. Minimap2 uses the same base algorithm for all
applications, but it has to apply different sets of parameters depending on
input data types. Similar to BWA-MEM, minimap2 introduces `presets' that
modify multiple parameters with a simple invocation. Detailed settings
and command-line options can be found in the minimap2 manpage. In addition to
the applications evaluated in the following sections, minimap2 also retains
minimap's functionality to find overlaps between long reads and to search
against large multi-species databases such as \emph{nt} from NCBI.

\subsection{Aligning long genomic reads}\label{sec:long-genomic}

\begin{figure}[!tb]
\centering
\includegraphics[width=.5\textwidth]{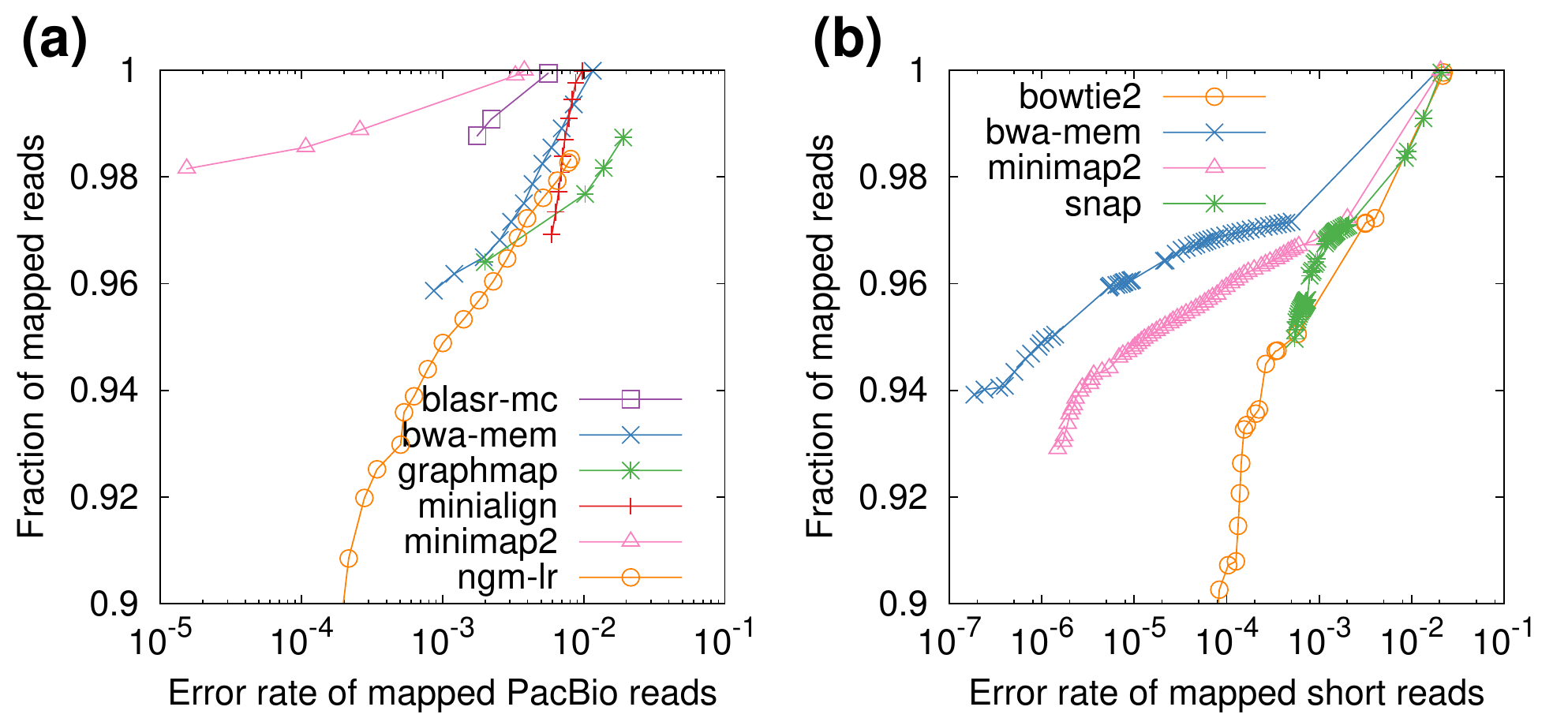}
\caption{Evaluation on aligning simulated reads. Simulated reads were mapped 
to the primary assembly of human genome GRCh38. A read is considered correctly
mapped if its longest alignment overlaps with the true interval, and the
overlap length is $\ge$10\% of the true interval length.  Read alignments are
sorted by mapping quality in the descending order. For each mapping quality
threshold, the fraction of alignments (out of the number of input reads) with
mapping quality above the threshold and their error rate are
plotted along the curve. (a) long-read alignment evaluation. 33,088 $\ge$1000bp
reads were simulated using pbsim~\citep{Ono:2013aa} with error profile sampled
from file `m131017\_060208\_42213\_*.1.*' downloaded at
\href{http://bit.ly/chm1p5c3}{http://bit.ly/chm1p5c3}. The N50 read length is
11,628. Aligners were run under the default setting for SMRT reads.
Kart outputted all alignments at mapping quality 60, so is not shown in the
figure. It mapped nearly all reads with 4.1\% of alignments being wrong, less
accurate than others.  (b) short-read alignment evaluation. 10 million pairs of
150bp reads were simulated using mason2~\citep{Holtgrewe:2010aa} with option
`\mbox{--illumina-prob-mismatch-scale 2.5}'. Short-read aligners were run under
the default setting except for changing the maximum fragment length to
800bp.}\label{fig:eval}
\end{figure}

As a sanity check, we evaluated minimap2 on simulated human reads along with
BLASR~(v1.MC.rc64; \citealp{Chaisson:2012aa}),
BWA-MEM~(v0.7.15; \citealp{Li:2013aa}),
GraphMap~(v0.5.2; \citealp{Sovic:2016aa}),
Kart~(v2.2.5; \citealp{Lin:2017aa}),
minialign~(v0.5.3; \href{https://github.com/ocxtal/minialign}{https://github.com/ocxtal/minialign}) and
NGMLR~(v0.2.5; \citealp{Sedlazeck169557}). We excluded rHAT~\citep{Liu:2016ab}
and LAMSA~\citep{Liu:2017aa} because they either
crashed or produced malformatted output. In this evaluation, minimap2 has
higher power to distinguish unique and repetitive hits, and achieves overall
higher mapping accuracy (Fig.~\ref{fig:eval}a). Minimap2 and
NGMLR provide better mapping quality estimate: they rarely give repetitive hits
high mapping quality.  Apparently, other aligners may
occasionally miss close suboptimal hits and be overconfident in wrong mappings.
On run time, minimap2 took 200 CPU seconds, comparable to minialign and Kart, and is over
30 times faster than the rest.  Minimap2 consumed 6.8GB memory at the peak,
more than BWA-MEM (5.4GB), similar to NGMLR and less than others.

On real human SMRT reads, the relative performance and fraction of mapped reads reported by
these aligners are broadly similar to the metrics on simulated data. We are
unable to provide a good estimate of mapping error rate due to the lack of the
truth.  On ONT $\sim$100kb human reads~\citep{Jain128835}, BWA-MEM failed.
Kart, minialign and minimap2 are over 70 times faster than others. We have also
examined tens of $\ge$100bp INDELs in IGV~\citep{Robinson:2011aa} and can
confirm the observation by~\citet{Sedlazeck169557} that BWA-MEM often breaks
them into shorter gaps. The issue is much alleviated with minimap2, thanks
to the 2-piece affine gap cost.

\subsection{Aligning long spliced reads}

We evaluated minimap2 on SIRV control data~(AC:SRR5286959;
\citealp{Byrne:2017aa}) where the truth is known. Minimap2 predicted 59\,918
introns from 11\,018 reads. 93.8\% of splice juctions are precise. We examined
wrongly predicted junctions and found the majority were caused by clustered
splicing signals (e.g. two adjacent ${\tt GT}$ sites). When INDEL sequencing
errors are frequent, it is difficult to find precise splicing sites in this
case. If we allow up to 10bp distance from true splicing sites, 98.4\% of
aligned introns are approximately correct. It is worth noting that for SIRV, we
asked minimap2 to model the ${\tt GT..AG}$ splicing signal only without extra
bases. This is because SIRV does not honor the evolutionarily prevalent signal
${\tt GT[A/G]..[C/T]AG}$~\citep{Irimia:2008aa}.

\begin{table}[!tb]
\processtable{Evaluation of junction accuracy on 2D ONT reads}
{\footnotesize\label{tab:intron}
\begin{tabular}{p{3.1cm}rrrr}
\toprule
& GMAP & minimap2 & SpAln & STAR\\
\midrule
Run time (CPU min)        & 631      & 15.9     & 2\,076   & 33.9 \\
Peak RAM (GByte)          & 8.9      & 14.5     & 3.2      & 29.2\vspace{1em}\\
\# aligned reads          & 103\,669 & 104\,199 & 103\,711 & 26\,479 \\
\# chimeric alignments    & 1\,904   & 1\,488   & 0        & 0 \\
\# non-spliced alignments & 15\,854  & 14\,798  & 17\,033  & 10\,545\vspace{1em}\\
\# aligned introns        & 692\,275 & 693\,553 & 692\,945 & 78\,603 \\
\# novel introns          & 11\,239  & 3\,113   & 8\,550   & 1\,214 \\
\% exact introns          & 83.8\%   & 94.0\%   & 87.9\%   & 55.2\% \\
\% approx. introns        & 91.8\%   & 96.9\%   & 92.5\%   & 82.4\% \\
\botrule
\end{tabular}
}{Mouse cDNA reads (AC:SRR5286960; R9.4 chemistry) were mapped to the primary assembly of mouse
genome GRCm38 with the following tools and command options: minimap2 (`-ax
splice'); GMAP (`-n 0 --min-intronlength 30 --cross-species'); SpAln (`-Q7 -LS
-S3'); STARlong (according to
\href{http://bit.ly/star-pb}{http://bit.ly/star-pb}). The alignments were
compared to the EnsEMBL gene annotation, release 89. A predicted intron
is \emph{novel} if it has no overlaps with any annotated introns. An intron
is \emph{exact} if it is identical to an annotated intron. An intron is
\emph{approximate} if both its 5'- and 3'-end are within 10bp around the ends
of an annotated intron. Chimeric alignments are defined in the SAM spec~\citep{Li:2009ys}.}
\end{table}

We next aligned real mouse reads~\citep{Byrne:2017aa} with GMAP~(v2017-06-20;
\citealp{Wu:2005vn}), minimap2, SpAln~(v2.3.1; \citealp{Iwata:2012aa}) and
STAR~(v2.5.3a; \citealp{Dobin:2013kx}). In general, minimap2 is more
consistent with existing annotations (Table~\ref{tab:intron}): it finds
more junctions with a higher percentage being exactly or approximately correct.
Minimap2 is over 40 times faster than GMAP and SpAln. While STAR is close to
minimap2 in speed, it does not work well with noisy reads.

We have also evaluated spliced aligners on a human Nanopore Direct RNA-seq
dataset (\href{http://bit.ly/na12878ont}{http://bit.ly/na12878ont}). Minimap2
aligned 10 million reads in $<$1 wall-clock hour using 16 CPU cores. 94.2\% of
aligned splice junctions consistent with gene annotations. In comparison,
GMAP under option `-k 14 -n 0 --min-intronlength 30 --cross-species' is 160
times slower; 68.7\% of GMAP junctions are found in known gene annotations. The
percentage increases to 84.1\% if an aligned junction within 10bp from an
annotated junction is considered to be correct. On a public Iso-Seq dataset
(human Alzheimer brain from
\href{http://bit.ly/isoseqpub}{http://bit.ly/isoseqpub}), minimap2 is also
faster at higher junction accuracy in comparison to other aligners in
Table~\ref{tab:intron}.

We noted that GMAP and SpAln have not been optimized for noisy reads. We are
showing the best setting we have experimented, but their developers should be
able to improve their accuracy further.


\subsection{Aligning short genomic reads}

We evaluated minimap2 along with Bowtie2~(v2.3.3; \citealt{Langmead:2012fk}), BWA-MEM and
SNAP (v1.0beta23; \citealt{Zaharia:2011aa}). Minimap2 is 3--4 times as fast as Bowtie2 and
BWA-MEM, but is 1.3 times slower than SNAP. Minimap2 is more accurate on this
simulated data set than Bowtie2 and SNAP but less accurate than BWA-MEM
(Fig.~\ref{fig:eval}b). Closer investigation reveals that BWA-MEM achieves
a higher accuracy partly because it tries to locally align a read in a small
region close to its mate. If we disable this feature, BWA-MEM becomes slightly
less accurate than minimap2. We might implement a similar heuristic
in minimap2 in future.

To evaluate the accuracy of minimap2 on real data, we aligned human reads
(AC:ERR1341796) with BWA-MEM and minimap2, and called SNPs and small INDELs
with GATK HaplotypeCaller v3.5~\citep{Depristo:2011vn}. This run was sequenced
from experimentally mixed CHM1 and CHM13 cell lines. Both of them are homozygous
across the whole genome and have been \emph{de novo} assembled with SMRT reads
to high quality. This allowed us to construct an independent truth variant
dataset~\citep{Li223297} for
ERR1341796. In this evaluation, minimap2 has higher SNP false negative rate
(FNR; 2.6\% of minimap2 vs 2.3\% of BWA-MEM), but fewer false positive SNPs per
million bases (FPPM; 7.0 vs 8.8), similar INDEL FNR (11.2\% vs 11.3\%) and
similar INDEL FPPM (6.4 vs 6.5). Minimap2 is broadly comparable to BWA-MEM in the
context of small variant calling.

\subsection{Aligning long-read assemblies}

Minimap2 can align a SMRT assembly (AC:GCA\_001297185.1) against GRCh38 in 7
minutes using 8 CPU cores, over 20 times faster than nucmer from
MUMmer4~\citep{Marcais:2018aa}. With the paftools.js script from the minimap2
package, we called 2.67 million single-base substitutions out of 2.78Gbp
genomic regions. The transition-to-transversion ratio (ts/tv) is 2.01. In
comparison, using MUMmer4's dnadiff pipeline, we called 2.86 million
substitutions in 2.83Gbp at ts/tv=1.87. Given that ts/tv averaged across the
human genome is about 2 but ts/tv averaged over random errors is 0.5, the
minimap2 callset arguably has higher precision at lower sensitivity.

The sample being assembled is a female. Minimap2 still called 201 substitutions
on the Y chromosome. These substitutions all come from one contig aligned at
96.8\% sequence identity. The contig could be a segmental duplication
absent from GRCh38. In constrast, dnadiff called 9070 substitutions on the Y
chromosome across 73 SMRT contigs. This again implies our minimap2-based
pipeline has higher precision.

\section{Discussions}

Minimap2 is a versatile mapper and pairwise aligner for nucleotide sequences.
It works with short reads, assembly contigs and long noisy genomic and RNA-seq
reads, and can be used as a read mapper, long-read overlapper or a full-genome
aligner. Minimap2 is also accurate and efficient, often outperforming other
domain-specific alignment tools in terms of both speed and accuracy.

The capability of minimap2 comes from a fast base-level alignment algorithm and
an accurate chaining algorithm. When aligning long query sequences, base-level
alignment is often the performance bottleneck. The Suzuki-Kasahara algorithm
greatly alleviates the bottleneck and enables DP-based splice alignment
involving $>$100kb introns, which was impractically slow ten years ago.  The
minimap2 chaining algorithm is fast and highly accurate by itself.  In fact,
chaining alone is more accurate than all the other long-read mappers in
Fig.~\ref{fig:eval}a (data not shown). This accuracy helps to reduce downstream
base-level alignment of candidate chains, which is still several times slower than
chaining even with the Suzuki-Kasahara improvement. In addition, taking a
general form, minimap2 chaining can be adapted to non-typical data types such as
spliced reads and multiple reads per fragment. This gives us the opportunity to
extend the same base algorithm to a variety of use cases.

Modern mainstream aligners often use a full-text index, such as suffix array or
FM-index, to index reference sequences. An advantage of this approach is that
we can use exact seeds of arbitrary lengths, which helps to increase seed
uniqueness and reduce unsuccessful extensions. Minimap2 indexes reference
k-mers with a hash table instead. Such fixed-length seeds are inferior to
variable-length seeds in theory, but can be computed much more efficiently in
practice. When a query sequence has multiple seed hits, we can afford to skip
highly repetitive seeds without affecting the final accuracy. This further
alleviates the concern with the seeding uniqueness. At the same time, at low
sequence identity, it is rare to see long seeds anyway. Hash table is the ideal
data structure for mapping long noisy sequences.

\section*{Acknowledgements}
We owe a debt of gratitude to H. Suzuki and M. Kasahara for releasing their
masterpiece and insightful notes before formal publication. We thank M.
Schatz, P. Rescheneder and F. Sedlazeck for pointing out the limitation of
BWA-MEM. We are also grateful to minimap2 users who have greatly helped to
suggest features and to fix various issues.

\paragraph{Funding\textcolon} NHGRI 1R01HG010040-01

\bibliography{minimap2}

\end{document}